# Physical meaning and derivation of the Schrodinger and Dirac equations

Spyros Efthimiades
*Department of Natural Sciences, Fordham University, New York, NY 10023*
*email*: sefthimiades@fordham.edu

The wavefunction of a particle is obtained from its intermediate states and interaction processes considered as happening concurrently. When the interaction is described by a potential, the energy of the particle is equal to its total kinetic plus potential energies. We derive the Schrodinger and Dirac equations as the unique conditions the wavefunction must satisfy at each point in order to fulfill the corresponding energy equation.

PACS numbers: 03.65.−w.

Keywords: Schrodinger equation; Dirac equation; Quantum dynamics.

Non-relativistic quantum mechanics has been based on the postulated Schrodinger equation that "plays a role analogous to Newton's second law" and which "cannot be derived from simpler principles but it is the fundamental principle of quantum mechanics".[1-2] Multiplying the Schrodinger equation by $\psi^*$ and integrating over space and time we get the energy equation but it is unclear that the reverse would hold. However, logic and physical perception suggest that the two equations are uniquely interconnected. We will derive that the Schrodinger equation is the unique condition the wavefunction must satisfy at each point in order to fulfill the energy equation.

A particle carrying energy $E$ and interacting with a time-independent potential $V(\mathbf{r})$ has specific total potential energy ($PE$) and, consequently, specific total kinetic energy ($KE$). The wavefunction of the particle is the superposition of its intermediate wave states that all have energy $E$ and so the time-energy dependence of the wavefunction can be extracted as an unobservable phase factor. The space part $\psi(\mathbf{r})$ of the wavefunction is a superposition of particle waves having amplitudes $a(\mathbf{p})$.

$$\psi(\mathbf{r}) = \frac{1}{(2\pi\hbar)^{3/2}} \int a(\mathbf{p}) e^{\frac{i}{\hbar} p \cdot r} d^3\mathbf{p} \tag{1}$$

The description of the interaction by a potential introduces the requirement that the wavefunction fulfills the energy equation

$$E = KE + PE \qquad \rightarrow \qquad E = \int |a(\mathbf{p})|^2 \frac{\mathbf{p}^2}{2m} d^3\mathbf{p} + \int |\psi(\mathbf{r})|^2 V(\mathbf{r}) d^3\mathbf{r} \tag{2}$$

Since $\psi(\mathbf{r})$ and $a(\mathbf{p})$ are interconnected through equation (1), the potential and kinetic energies are interrelated and the above energy equation is fulfilled by a specific wavefunction only.

In particle scatterings, the particle energy ($E$) and the interaction potential $V(\mathbf{r})$ determine the wavefunction $\psi(\mathbf{r})$ − the unique superposition of wave-states for which the kinetic plus potential energies equal $E$. For bound particles, the potential $V(\mathbf{r})$ plus boundary conditions determine the possible eigenfunctions $\psi_n(\mathbf{r})$ that fulfill the energy equation and yield the energy eigenvalues $E_n$.

From (1) we have

$$\int |a(\mathbf{p})|^2 \frac{\mathbf{p}^2}{2m} d^3\mathbf{p} = \int \psi^* \left[ -\frac{\hbar^2}{2m} \nabla^2 \psi \right] d^3\mathbf{r} \tag{3}$$

and substituting (3) into (2), we put the energy equation in the following form

$$\int \psi^* \left[ E\psi - \frac{\hbar^2}{2m} \nabla^2 \psi - V(\mathbf{r})\psi \right] d^3\mathbf{r} = 0 \tag{4}$$

The above equation states that the average value of the integrand is zero. But, as we will justify below, the integrand itself is zero at each point and, dividing it by $\psi^*$, we get

$$-\frac{\hbar^2}{2m} \nabla^2 \psi + V(\mathbf{r})\psi = E\psi \qquad \textit{at each } \mathbf{r} \tag{5}$$

This is the Schrodinger equation, derived as the unique condition the wavefunction must satisfy in order to fulfill the energy equation. Only a wavefunction that satisfies this condition produces the kinetic and potential energies that fulfill the energy equation.

The integrand $I(\mathbf{r})$ in (4) must vanish everywhere because if it differs from zero at some point this would alter the value of the potential. For example, suppose that the integrand at point $\mathbf{r}$ is $I(\mathbf{r}) = N(\mathbf{r}) = \psi^* n(\mathbf{r})\psi$. Then, we would have

$$\begin{aligned} I(\mathbf{r}) &= \psi^* \left[ E\psi + \frac{\hbar^2}{2m} \nabla^2 \psi - V(\mathbf{r})\psi \right] = \psi^* n(\mathbf{r})\psi \\ I'(\mathbf{r}) &= \psi^* \left[ E\psi + \frac{\hbar^2}{2m} \nabla^2 \psi - \{V(\mathbf{r}) + n(\mathbf{r})\}\psi \right] = 0 \end{aligned} \tag{6}$$

The value of the potential changes, arbitrarily, from $V(\mathbf{r})$ to $V(\mathbf{r})+n(\mathbf{r})$ and the wavefunction changes too. The amplitudes of the intermediate states of the new wavefunction depend on $V(\mathbf{r})$ and $n(\mathbf{r})$ – a local change of the potential generates a global change in the wavefunction. We get the wavefunction for the given potential for $n(\mathbf{r})=0$, therefore the Schrodinger equation [$I(\mathbf{r})=0$] is the unique physical condition that satisfies equation (6).

A classical potential is a strict but effective description of the interaction. Even though quantum energy fluctuations do happen, the potential is the average over these fluctuations. We may say that, in the non-relativistic theory, the particle "sees" only the classical potential and the superposition of its intermediate states "sculpts" the wavefunction that yields the proper potential and kinetic energies.

To provide a concrete example of the concepts relevant to the derivation of the Schrodinger equation, we consider the ground state of the electron in a Hydrogen atom.

Experimentally, we can detect the intermediate electron momentum states and measure their amplitudes by scattering energetic photons off Hydrogen atoms. Such "photoelectric" experiments have been carried out with Hydrogen and Helium atoms.[3] From these measurements, we determine that the probability of finding the electron in an intermediate momentum state $p$ is

$$momentum\ \ probability = \left|a_1(p)\right|^2 = \frac{1}{\pi^2}\left(\frac{2r_0}{\hbar}\right)^3\left[1+\frac{r_0^2}{\hbar^2}p^2\right]^{-4} \tag{7}$$

where $r_0 = 0.053$ nm.

Inserting $a_1(p)$ into (1) we obtain the eigenfunction $\psi_1(r)$ and we get that the probability of finding the electron at distance $r$ from the nucleus is

$$position\ \ probability = \left|\psi_1(r)\right|^2 = \left[\frac{1}{\sqrt{\pi r_0^3}}e^{-\frac{r}{r_0}}\right]^2 \tag{8}$$

All intermediate electron waves on the ground level have energy $E_1$, equal to the total kinetic plus potential energies that we calculate from $|a_1(p)|^2$ and $|\psi_1(r)|^2$.

$$E_1 = \int \left|a_1(p)\right|^2 \frac{\mathbf{p}^2}{2m}d^3\mathbf{p} + \int \left|\psi_1(r)\right|^2 \frac{-ke^2}{r}d^3\mathbf{r} = +13.6 - 27.2 = -13.6\ eV \tag{9}$$

Theoretically, we can obtain $\psi_1(r)$ and $a_1(p)$ as follows: An electron eigenstate is a superposition of intermediate wave-states having energy $E_n$. Since there are no boundaries, intermediate states of all momenta (from $-\infty$ to $+\infty$) arise in every direction. An electron eigenfunction $\psi_n(\mathbf{r})$ must vanish at infinity and, in order to fulfill the energy relation, must satisfy at each point $\mathbf{r}$ the Schrodinger equation

$$-\frac{\hbar^2}{2m}\nabla^2\psi_n - \frac{ke^2}{r}\psi_n = E_n\psi_n \tag{10}$$

For the ground electron eigenstate we have

$$\begin{aligned}\psi_1(\mathbf{r},t) &= \psi_1(\mathbf{r}) \otimes e^{-\frac{i}{\hbar}E_1 t} \\ \psi_1(\mathbf{r}) &= \int_{-\infty}^{+\infty} a_1(\mathbf{p})e^{\frac{i}{\hbar}p\cdot r}d^3\mathbf{p}\end{aligned} \tag{11}$$

Solving equation (10) we obtain $\psi_1(\mathbf{r})$ and then calculate $a_1(\mathbf{p})$. These calculated results agree with the experimental ones shown in (7) and (8), where now we get $r_0 = \hbar^2/(mke^2) = 0.053$ nm and $E_1 = -ke^2/(2r_0) = -13.6$ eV.

When the potential $V(\mathbf{r},t)$ depends on space *and* time, the wavefunction $\psi(\mathbf{r},t)$ is a superposition of intermediate waves having different momenta and energies.

$$\psi(\mathbf{r},t) = \frac{1}{(2\pi\hbar)^{4/2}} \int a(\mathbf{p},E)\, e^{\frac{i}{\hbar}(p\cdot r - Et)} d^3\mathbf{p}dE \qquad (12)$$

Following steps similar to those used for deriving the time-independent equation, we derive from the energy equation the time-dependent Schrodinger equation

$$E = KE + PE \qquad \rightarrow \qquad i\hbar\frac{\partial\psi(\mathbf{r},t)}{\partial t} = -\frac{\hbar^2}{2m}\nabla^2\psi(\mathbf{r},t) + V(\mathbf{r},t)\psi(\mathbf{r},t) \qquad (13)$$

The relativistic energy equation for a (spin 1/2) electron inside an electromagnetic field $A_\mu(A_0,A_i)$ can be put in the linearized form

$$\gamma_0\left(E - eA_0\right) = \gamma_i\left(p_i - \frac{e}{c}A_i\right)c + mc^2 \qquad (14)$$

where $E$, $A_0$, $p_i$ and $A_i$ represent total values and $\gamma_0$, $\gamma_i$ are the Dirac matrices.

Proceeding in steps analogous to those used for deriving the Schrodinger equations, we get the following wavefunction equation

$$\gamma_0\left(i\hbar\frac{\partial}{\partial t} - eA_0\right)\psi(x) = \gamma_i\left(-i\hbar\frac{\partial}{\partial x_i} - \frac{e}{c}A_i\right)c\,\psi(x) + mc^2\psi(x) \qquad \textit{at each } x \qquad (15)$$

where $\psi(x)$ is a superposition of spin wave states having different momenta and energies. This is the Dirac equation that the electron wavefunction must satisfy at each space-time point $x(t,\mathbf{x})$ in order to fulfill the relativistic energy equation.

The derivation of the Schrodinger equation removes its axiomatic status and we can unify quantum dynamics under one principle: The probabilities of the physical outcomes are obtained from the intermediate waves and interaction processes considered as happening concurrently. The intermediate waves may arise: from boundaries (i.e. double slit experiment, light reflection, light refraction, tunneling, particle in a box); from scattering events contributing to a certain outcome (i.e. electron-nucleus collision); from the superposition of wave-states producing the total (potential plus kinetic) energy of particles (i.e. atomic electron eigenfunctions); from the addition of quantum interactions processes (i.e. Compton scattering in quantum electrodynamics), etc. A more complete description of the principles and dynamics of quantum mechanics can be found in Reference (4).

The derivations of the wavefunction equations of quantum mechanics lead to a theoretically consistent and experimentally justified quantum theory.